\documentclass{iau}
\usepackage{graphicx} 
\usepackage{upgreek}
\title[IAUS291.~~Radio pulsar variability] %% short title %%
{Radio pulsar variability} %% full title %%

\author[E.~F. Keane]  %% short author list %%
{E.~F. Keane
}

\affiliation{Max Planck Institut f\"{u}r Radioastronomie, \\ Auf
  dem H\"{u}gel 69, D-53121, Bonn, Germany. \\ email: {\tt
    ekeane@mpifr-bonn.mpg.de}}

\pubyear{2012}
\volume{291}  %% insert here IAU Symposium No.
\jname{\mbox{Neutron Stars and Pulsars: Challenges and Opportunities after 80 years}}
\editors{J. van Leeuwen, ed.} 
\begin{document}

\maketitle

%% -- Abstract ----------------------------------
\begin{abstract}

  Pulsars are potentially the most remarkable physical laboratories we
  will ever use. Although in many senses they are extremely clean
  systems there are a large number of instabilities and variabilities
  seen in the emission and rotation of pulsars. These need to be
  recognised in order to both fully understand the nature of pulsars,
  and to enable their use as precision tools for astrophysical
  investigations. Here I describe these effects, discuss the wide
  range of timescales involved, and consider the implications for
  precision pulsar timing.

\end{abstract}

% add below any authors, subjects and objects for indexing 
%   add more lines if necessary
%   but leave all lines commented out
%\index[author]{LastName1, Initials|textbf}
%\index[author]{LastName2, Initials|textbf}
%\index[subject]{Keyword1}
%\index[subject]{Keyword2}
%\index[object]{Object1}
%\index[object]{Object2}

\firstsection % if your document starts with a section,

\section{Introduction}\label{sec:intro}
A textbook pulsar emits a beam of radio emission from just above its
magnetic poles. The mis-alignment of the spin and magnetic axes then
results in a light-house effect as the star rotates. Those pulsars
whose radio beams cut across the Earth are observed as a string of
sharp pulses in the signals detected by radio telescopes. The signal
is straight forward to model with a simple slow-down law consisting
(usually) of just spin frequency, its derivative and (if applicable)
some binary parameters. The regularity of the signal means that these
pulses act as the ticks of an extremely precise clock. Furthermore,
pulsars are often to be found in extreme environments, which we are
able to study by utilising this clock-like nature. The moniker of
`super clocks in space' is well earned.

A real-life pulsar deviates from the ideal in a number of ways. This
is due to a number of instrumental, propagation and intrinsic effects,
many of which are not well understood. In \S~\ref{sec:what_do_we_see}
we discuss the wide range of variable behaviour observed in pulsar
signals. In \S~\ref{sec:how_do_they_work} we consider how pulsars
actually work before asking why this is of interest to pulsar
astronomers in \S~\ref{sec:who_cares}. Finally, in
\S~\ref{sec:conclusions_discussion}, we present conclusions and
discussions.

\section{What do we see?}\label{sec:what_do_we_see}
The range of variability timescales in pulsars is remarkably wide,
spanning all timescales on which it has been possible to measure. The
fastest timescales to have been probed are nanoseconds. The voltage
signals from radio telescopes are commonly Nyquist-sampled at rates of
$\sim1$~GHz, but usually this time resolution is traded for frequency
resolution, and furthermore data is integrated in time to increase the
signal-to-noise ratio (S/N). However, in the case of the Crab pulsar
this is not necessary in order to detect a signal, and \cite{hkwe03}
have observed kJy pulses lasting 2~ns, showing that its well-known
$\sim\upmu$s `giant pulses' are in fact composed of a large number of
such ``shots''. These shots appear to be the quanta of pulsar radio
emission. They are not resolved --- indeed, we might expect this,
i.e. an intrinsic timescale of $\lesssim 100$~ps, given the
uncertainty principle and the observations that pulsars emit over
bandwidths of several tens of GHz~\cite[(Maron \etal\  2000; Camilo \etal\  2007)]{mkk+00,crp+07}. The actual
mechanism is unknown but the brightness temperature of
$T_{\mathrm{B}}=10^{37}$~K (for the Crab pulses) implies, using the
well-known expression for the maximum possible brightness temperature
$T_{\mathrm{B,max}}=6\times 10^9 N(\gamma-1)$~K, a coherence factor of
$N\approx 10^{27}/\gamma$. Clearly the mechanism is coherent, most
likely involving particles emitting in bunches, a plasma instability
or some kind of maser, but despite much
effort~\cite[(Ginzberg \& Zheleznyakov 1970; Asseo \etal\  1990; Lyutikov \etal\  1999; Melrose 2004)]{gz70,aps90,lbm99,mel04b} the details are not known.
%involving a large number of particles within a ``chapati-shaped''
%coherence volume of $\sim \gamma^2\lambda^3/\pi$.

The duration of a time sample in most pulsar observations is usually
$\gg 100$~ps so that a
%much longer than the (presumed) intrinsic $\sim 100$~ps timescale of
%the emission quanta. A 
large number of shots are incoherently added within each time
sample. The Poisson distribution of the shots then approaches that of
a Gaussian, and it is common to model the pulsar signal
%, $y(t)$, as Gaussian noise, $n(t)$, modulated by an amplitude
%function, $a(t)$, i.e. $y(t) = n(t)a(t)$
as amplitude-modulated noise~\cite[(Rickett 1975)]{ric75}. This model is insufficient
however, as single pulse studies show non-Gaussian variations on
$\upmu$s$-$ms scales, e.g. the ``giant micropulses'' seen in Vela by
\cite{jvkb01}, and we see dramatic variations from one pulse period
to the next, on ms$-$s scales, 
%(i.e. $a(t)$ is not constant) 
e.g. we see changes in intensity (by factors of $\gtrsim 10^3$),
phase, pulse shape and the number of components.

Extremely organised behaviour is seen on second to minute timescales,
in the form of sub-pulse drifting. Here, a `Joy Division plot' reveals
that the pulses drift periodically (both earlier and later) in pulse
phase in regular `bands' as a function of time with typical repetition
periods of tens of spin periods. A standard explanation for this
behaviour has been the ``carousel model''~\cite[(Ruderman \& Sutherland 1975)]{rs75} where disparate
emission spots above the stellar surface are induced to rotate by $E
\times B$ drift. Lately however it has been shown that this model does
not explain the drifting seen in PSR~B0809$+$74 (Hassall et al., in
prep.). In a study of 187 pulsars, using the Westerbork Synthesis
Radio Telescope, \cite{wes06} showed that at least one third
exhibited drifting.

On timescales of seconds to minutes, and even up to hours we see
further organised behaviour in the form of `moding' --- the changing
of the pulse profile to one of a small number of different
profiles. If there is no detected radio emission from one of these
`modes' the phenomenon is commonly termed `nulling'. A quantitative
analysis of the pulse amplitude distributions of a large number of
pulsars has recently been performed by \cite{bjb+12}. This work
looked at the single pulse statistics of 315 pulsars with detectable
single pulses in the High Time Resolution Universe survey. The authors
classify the pulse amplitude distributions as either Gaussian (7
sources, 2\%), log-normal (84 sources, 27\%), multi-peaked (18
sources, 6\%) or unimodal (24 sources, 8\%). Unfortunately the
majority (182 sources, 58\%) did not fit within these
classifications. While we might suggest testing for more complex
distributions for the unclassified sources,
%, e.g. a log-normal distribution with a power-law tail as is seen in
%some pulsars such as PSR~B0656$+$14~\cite[(Weltevrede \etal\  2006)]{wsrw06}
this is not possible due to a paucity of detected pulses. Of the
unclassified 182 sources, only 92 had more than 20 detected pulses
during the 9-minute survey pointings, and a single pulse was all that
was detected for 22 of the sources.

Moding is also observed on timescales of hours to weeks, or even
months. The first realisation of this came when \cite{klo+06}
discovered that PSR~B1931$+$24 is detectable as a radio pulsar only
for periods of $\sim5-10$ days before `turning off' and remaining
undetectable for $\sim25-35$ days in a quasi-periodic
fashion. Crucially this moding is accompanied by a $\sim50\%$ change
in the spin-down rate, with
$\dot{\nu}_{\mathrm{hi}}/\dot{\nu}_{\mathrm{lo}}=1.5$. Since then two
more ``intermittent pulsars'' have been reported ---
PSRs~J1841$-$0500~\cite[(Camilo \etal\  2012)]{crc+12} and
J1832$+$0029~\cite[(Lorimer \etal\  2012)]{llm+12}. These sources have `on' and `off'
timescales $\sim 20-30$ times longer than B1931$+$24 and spin-down
rate ratios of $2.5$ and $1.8$ respectively. \cite{lhk+10} presented
results of several decades of Lovell Telescope observations of 17
pulsars where correlated quasi-periodic changes in $\dot{\nu}$ and
pulse profile were clearly observed. The changes in spin-down rate
ranged from $0.3\%$ to $13\%$. More examples of such behaviour
continue to be identified (see e.g. Karastergiou, these
proceedings). We stress that it is not simply switching between two
states that is seen in moding pulsars (see Fig.~5 from \cite{bb10} or
Fig.~1 from \cite{eamn12} for some excellent examples). Furthermore
we note that such moding is seen on all timescales ranging from
several years down to one rotation period~\cite[(Keane \& McLaughlin 2011)]{km11}. On the shorter
time-scales changes in $\dot{\nu}$ cannot be measured ---
\cite{ysw+12} point out that for modes persistent for less than a day
spin-down rate switching of a few percent would never be
detectable. The associated profile changes are also often quite subtle
and obviously cannot be discerned from pulse-to-pulse variations when
the persistence of the mode is less than the required duration to
surpass the stable profile threshold (see
\S~\ref{sec:who_cares}). \textit{It would seem that switching between
  a number of stable states, often with some quasi-periodicity, is a
  generic feature of pulsars}.
%The evidence points to pulsars not being stable either in their
%emission or their rotation\footnote{This is in addition to the effect
%of ``glitches''\cite[(Espinoza \etal\  2011)]{elsk11}, which themselves can result in
%persistent $\dot{\nu}$ switches and changes in
%emission~\cite[(Lyne \etal\  2009; Weltevrede \etal\  2011)]{lmk+09,wje11}.}.

Of course we must not forget that the emitted broadband signal from a
pulsar is subject to the transfer function of the interstellar medium
(which is also time variable on a number of scales) and that of the
telescope-receiver system itself (which will also have a number of
systematic contributions).
%, all of which must be considered in creating a model for the
%observed pulsar emission (see e.g. \cite{jg12,jgd12}).
There is an equally long list of these effects which must be accounted
for in modelling the pulsar signal but which I will not elaborate upon
here (but see e.g. \cite[Cordes \& Shannon 2010]{cs10}). Table~\ref{tab:PSR_timescales} gives
an incomplete list of timescales on which pulsars are known, or
expected, to be variable.

\begin{table}
  \begin{center}
    \caption{\small{An incomplete list of the variability and
        evolutionary timescales of a pulsar. A plethora of
        interstellar medium timescales also exist which will also
        modulate the observed pulsar signal, as well any gravitational
        wave sources. $\dagger$ Here we use the term `nulling', but
        `moding', `extreme pulse amplitude modulation' or a variety of
        similar terms could be used
        interchangeably.}}\label{tab:PSR_timescales}
    \vspace{10pt}
   \begin{tabular}{cll}
      \hline Timescale & Name & Cause \\
      \hline
      
      ns & Radio quanta ``shots'' & Fundamental emission timescale \\
      us$-$ms & Single pulse variations & ? \\
      ms$-$s & Pulse-to-pulse variations & ? \\
      s$-$min & Sub-pulse drifting & ? \\
      s$-$min & Nulling$\dagger$ & ? \\
      s$-$hrs & Extreme nulling & ? \\
      hrs$-$yrs & Quasi-periodic switching & Magnetospheric switching? \\
%      x & DISS & ISM \\
%      y & RISS & ISM \\
%      $\sim 1-10$~yr & gravitational wave (GW) contribution & stochastic GW background \\
      hrs$-$years & Orbital timescales & Orbital motion \\
      $\sim10^7$ years & NS cooling timescales & Thermal cooling \\
      $\sim 10^7$~yr & Galactic Evolution, $(G\rho_{\mathrm{MW}})^{-1/2}$ & Moving in Galactic potential\\
      $10^3-10^7$~yr & Spin Evolution, $P/\dot{P}$ & Loss of rotational energy \\

%      DM variations
%      desired residuals
%      pulse jitter
%      pulse stability timescale

      \hline
    \end{tabular}
  \end{center}
\end{table}

\section{How do they work?}\label{sec:how_do_they_work}
Assuming that the propagation and instrumental effects can be
understood (whether or not they can be removed) there are still a wide
range of transient behaviours seen in pulsars. This leads us to a big
question: \textit{How do we get erratic radio emission from a PSR with
  a particular timescale, and periodic switching?}

For force-free magnetospheres (see below) it has been shown that a
number of stable solutions exist with the closed magnetosphere not
necessarily extending to the light cylinder
radius~\cite[(Contopoulos \etal\  1999; Spitkovsky 2006)]{ckf99,spi06}. It has further been shown that perturbing
these solutions can result in a rapid switch from one magnetospheric
configuration to another~\cite[(Contopoulos 2005)]{con05}. However, these perturbations
are put in `by hand' and the underlying reason for the switching
remains unknown. Furthermore, why this would occur with a periodicity
is unknown. That the switching is quasi-periodic, rather than strictly
periodic, must also be explained. Recently \cite{sl12} have suggested
that the quasi-periodicity resembles that seen on ``the route to
chaos'' and detect chaotic behaviour in PSR~B1828$-$11, one of the
\cite{lhk+10} sample.
%\textit{How do we get erratic radio emission from a PSR with a
%  particular timescale \sout{and periodic switching}?}
The timescales for the erratic behaviour are wide (see
Table~\ref{tab:PSR_timescales}), so much so that it is difficult to
see what the decisive variables are.
%\textit{How do we get erratic radio emission from a PSR \sout{with a
%  particular timescale and periodic switching}?}
If the moding is simply a result of the magnetospheric
switching~\cite[(Timokhin 2010)]{tim10} the timescales for both phenomena are
obviously one and the same. This raises the question of whether
pulsars with large pulse-to-pulse modulation on much faster timescales
than the intermittent pulsars are changing magnetospheric
configuration constantly. This would suggest a picture of highly
unstable and frequent fast changes on the scale of the entire
magnetosphere. If this is not what is occurring in these cases it is
unclear on which timescales this ceases, as there seems to be a
continuum of moding/switching timescales observed~\cite[(Keane 2010a)]{kea10a}.
%\textit{How do we get \sout{erratic} radio emission from a PSR
%    \sout{with a particular timescale and periodic switching}?}
We are forced to abandon our big question entirely in favour of a more
tractable one: \textit{What does a PSR magnetosphere even look like?}
%\textit{\sout{How do we get erratic radio emission from a PSR with a
%    particular timescale and periodic switching?}}

There are two approaches to answering this question --- the first is
to solve Maxwell's equations for a rapidly-spinning strongly-magnetic
highly-conductive ball; the second is to try to determine the geometry
of the system from observations of the polarisation characteristics of
pulsar emission. Both of these approaches should result in the same
answer, but both are fraught with many difficulties. Here I briefly
describe the first approach, but refer the reader to the works of
Radhakrishnan, Cooke, Kramer, Karastergiou, Johnston, Weltevrede,
Rankin, Wright and Noutsos for information on the geometrical
approach. When calculating Maxwell's equations in the vicinity of the
neutron star it is found that there are trapping surfaces for charges
of opposite sign above the poles, and in the equatorial
plane. Particles get ripped from the stellar surface and are simply
trapped in these `electrospheres' with no pulsar-like behaviour (see
e.g. Fig.\ 2.4, \cite[Keane 2010b]{kea10b}). One then would assume that either the
initial conditions do not represent reality, i.e. in the violent
supernova explosion wherein the neutron star was born there was
abundant plasma provided from the offset so that the electrosphere
scenario never arises, or, that the electrosphere solution is in fact
unstable (e.g. to the diochotron instability, see Spitkovsky, these
proceedings) and breaks down after some time. Regardless of the reason
some authors have pressed on assuming ``a sufficiently large charge
density whose origin we do not question''~\cite[(Contopoulos \etal\  1999)]{ckf99} and solved
``the pulsar equation''~\cite[(Michel 1973)]{mic73} for the first time. The results
of this work show current flows in the magnetosphere coincident with
the `gap regions' for emission derived by the geometric approach so
that it seems that progress is being made towards understanding pulsar
magnetospheres. Table~\ref{tab:known_unknown} summarises some of the
knowns and unknowns.

\begin{table}
  \begin{center}
    \caption{\small{Some of the important questions regarding pulsar
        magnetospheres and the status of the force-free solutions (see
        e.g. Li; Spitkovsky, these
        proceedings).}}\label{tab:known_unknown}
    \begin{tabular}{ll}
      \hline Question & Status \\
      \hline
      
      Stable magnetosphere with dE/dt$>0$? & Yes \\
      Why force-free? & Don't Know \\
      $2+$ stable solutions possible? & Yes \\
      Switching between configurations? & Mechanism unknown \\
      Switching with (quasi-)periodicity? & No \\
      Braking index predictions? & Many ($n\neq3$) \\
      Radio emission explained? & No \\
      Gamma-ray emission explained? & Realistic lightcurves \\

      \hline
    \end{tabular}
  \end{center}
\end{table}

\section{Who cares?}\label{sec:who_cares}

\vspace{10pt}
\begin{quote}
%  \begin{singlespace}
    ``I don't care, I just want to do timing.'' Anonymous.
%  \end{singlespace}
\end{quote}
\vspace{10pt}

Some astronomers may not be very concerned with how pulsars actually
work, and only interested in pulsars for their use as clocks, e.g. to
use in pulsar timing arrays (PTAs). In this case the only question
that matters is whether or not pulsar profiles are stable for typical
PTA observations. Fortunately this can be measured, and one such
method involves calculating $\rho$, the cross-correlation coefficient
of the observed pulse profile with a template profile. If $1-\rho
\propto N^{-1}$, where $N$ is the number of periods folded into the
observed profile, then the profile is stable~\cite[(Liu \etal\  2012)]{lkl+12}. Longer
integrations improve the profile's S/N only and not its stability.
While the value of $N$ where the exponent transitions to $-1$ denotes
the stability timescale, different exponents reveal other timescales
at work, e.g. nulling/moding timescales if
present~\cite[(Keane 2010b)]{kea10b}. Although the stability of pulsar profiles is
implicitly assumed\footnote{It is assumed that the observed profile is
  a shifted scaled version of a smooth (sometimes analytic) template
  with additive noise.} in pulsar timing,
%and therefore is required to (say) use pulsars to detect
%gravitational waves, except for the
it is not clear whether this has been systematically confirmed for all
PTA pulsars.
%The only such work seems to be of
%`slow' pulsars in the 1970s, which are not stable for typical
%observation lengths and in any event not of interest for PTA work. 
The received wisdom is that $10^4$ periods gives you a stable profile
but \cite{lkl+12} found this to be dependent upon the pulsar with
values of up to $10^5$ periods required in some cases. 
%Likewise it
%could be that some pulsars are particularly stable much earlier than
%this but this has not been checked. The author encourages all pulsar
%timers to check if their favourite pulsar is stable for the typical
%observation times they use --- it is crucial to realise that high
It is important to note that a high S/N does not imply stability
(based solely on S/N we can time pulsars using their single pulses,
but this is not \textit{precision} pulsar timing, see \cite{kkl+11}
for details).
%The stability timescale also gives a natural limit on the resolution
%for resolving orbital features, e.g. if a pulsar had a Shapiro delay
%peak lasting $10^3$ periods (as in the double pulsar) and a profile
%which is stable only after $10^4$ periods then we must coherently add
%10 orbits (after iteratively determining a timing solution) to
%resolve the feature correctly. In this case using $10$ separate
%$10^3$ period observations is not equivalent. The difference may in
%fact be slight but such systematics must be considered when one is
%interested in resolving $10-100$~ns signals, over several years, in
%pulsar timing data (see Hobbs, these proceedings).

If one used pulsar profiles which were not stable then there would be
no justification for expecting a good fit to the timing model, with
$\chi^2_{\mathrm{red}}=1$. Oddly enough there is a practice (which is
admittedly dying out) to assume that the best fit model, i.e. the one
with the lowest $\chi^2_{\mathrm{red}}$ value, is the correct model,
and to then scale the errors so as to make
$\chi^2_{\mathrm{red}}=1$. The errors in this case are scaled by an
`EFAC' quantity. This is very bad practice for several reasons (see
\S~3.2.1 of \cite{and10} for more details), e.g. it assumes that: the
error distribution is Gaussian; the model is linear in all of its
parameters; the model used is correct (also completely negating the
point of using the chi-squared \textit{test}).

Pulse jitter is another contribution to errors in pulse
time-of-arrival measurements which is usually ignored. Jitter is only
evident in pulsar profiles when the S/N of single pulses are $\gtrsim
1$. Currently, for PTA sources, this is only relevant for
PSR~J0437$-$4715. For SKA-era sensitivity this must be accounted for
in all PTA pulsars, but fortunately this is possible, as has been
demonstrated for J0437$-$4715~\cite[(Liu \etal\  2012)]{lkl+12}.

\section{Conclusions \& Discussion}\label{sec:conclusions_discussion}
Pulsar emission and rotation is variable on a wide range of
timescales. It is vital to gain a full understanding of these things
in order to (a) understand pulsars; and (b) perform precision pulsar
timing. The author's bias suggests to him that it may be difficult to
achieve the latter with first making significant inroads into
achieving the former. For example the observed behaviour (described in
\S~\ref{sec:what_do_we_see}) suggest a number of questions which the
pulsar timing community should be thinking about: Is there any reason
why there would not be (perhaps periodic or quasi-periodic) spin-down
rate switching occurring in many/all pulsars? Is there any reason why
there would not be (perhaps periodic or quasi-periodic) spin-down rate
switching in many/all millisecond pulsars? Are there other (perhaps
deterministic) timing instabilities yet to be identified? The planned
upcoming studies of large pulsar timing databases (S. Johnston,
private communication) will no doubt shed valuable light on what the
answers to these questions are, and bring us a few steps closer to
understanding those super clocks in space.

\section*{Acknowledgements}
EK would like to thank the SOC for the invitation to speak at the
General Assembly, and the LOC for their hospitality throughout the
conference. EK is grateful 
to Lijing Shao for pointing out a very helpful reference paper, and to
Mark Purver for valuable comments on this text.

% CUP work flow only accepts EPS -- not PDF, JPG, etc.
% \begin{figure}[b]
% \begin{center}
%  \includegraphics[width=3.4in]{YourFig.eps} 
%  \caption{Path of pre-solar grains from their stellar sources to the
%    laboratory.} 
%    \label{fig1}
% \end{center}
% \end{figure}


\begin{thebibliography}{}

%\bibitem[A \etal\ (1995)]{A95}
%{A, A., B, B., \& C, C.} 1995,
%\textit{Meteoritics}, 1, 9

\bibitem[Andrae (2010)]{and10}
{Andrae, R.} 2010,
``{Error estimation in astronomy: A guide}'', astro-ph/1009.2755.

\bibitem[Asseo \etal\  (1990)]{aps90}
{Asseo, E., Pelletier, G. \& Sol, H.} 1990,
\textit{MNRAS}, 247, 529.

\bibitem[Burke-Spolaor \& Bailes (2010)]{bb10}
{Burke-Spolaor, S. \& Bailes, M.} 2010,
\textit{MNRAS}, 402, 855.

%\bibitem[Burke-Spolaor \etal\  (2011)]{bbj+11}
%{Burke-Spolaor, S. \etal\ } 2011,
%\textit{MNRAS}, 416, 2465.

\bibitem[Burke-Spolaor \etal\  (2012)]{bjb+12}
{Burke-Spolaor, S. \etal\ } 2012
\textit{MNRAS}, 423, 1351.

\bibitem[Camilo \etal\  (2007)]{crp+07}
{Camilo, F. \etal\ } 2007,
%{Camilo, F., Ransom, S.~M., Pe\~{n}alver, J., Karastergiou, A., van Kerkwijk, M.~H., Durant, M., Halpern, J.~P., Reynolds, J., Thum, C., Helfand, D.~J., Zimmerman, N., Cognard, I.} 2007,
\textit{ApJ}, 669, 561.

\bibitem[Camilo \etal\  (2012)]{crc+12}
%{Camilo, F., Ransom, S.~M., Chatterjee, S., Johnston, S. \& Demorest, P.} 2012,
{Camilo, F. \etal\ } 2012,
\textit{ApJ}, 746, 63.

%\bibitem[Chukwude \& Buchner (2012)]{cb12} 
%{Chukwude, A.~E. \& Buchner, S.} 2012
%\textit{ApJ}, 745, 40.

\bibitem[Contopoulos \etal\  (1999)]{ckf99}
{Contopoulos, I., Kazanas, D. \& Fendt, C.} 1999,
\textit{ApJ}, 511, 351.

\bibitem[Contopoulos (2005)]{con05}
{Contopoulos, I.} 2005,
\textit{A\&A}, 442, 579.

%\bibitem[Cordes \& Lazio (2002)]{cl02}
%{Cordes, J.~M. \& Lazio, T.~.J.~W.} 2002,
%astro-ph/0207156.

\bibitem[Cordes \& Shannon (2010)]{cs10}
{Cordes, J.~M. \& Shannon, R.~M.} 2010,
\textit{ApJ, submitted}, astro-ph/1010.3785.

\bibitem[Esamdin \etal\  (2012)]{eamn12}
%{Esamdin, A., Abdurixit, D., Manchester, R.~N. \& Niu, H.~B.} 2012,
{Esamdin, A. \etal\ } 2012,
\textit{ApJ}, 759, L3.

%\bibitem[Espinoza \etal\  (2011)]{elsk11}
%%{Espinoza, C.~M., Lyne, A.~G., Stappers, B.~W. \& Kramer, M.} 2011,
%{Espinoza, C.~M. \etal\ } 2011,
%\textit{MNRAS}, 414, 1679.

\bibitem[Ginzberg \& Zheleznyakov (1970)]{gz70}
{Ginzburg, V. L. \& Zheleznyakov, V. V.} 1970,
\textit{Comm. Astrophys.}, 2, 197.

\bibitem[Hankins \etal\  (2003)]{hkwe03}
%{Hankins, T.~H., Kern, J.~S., Weatherall, J.~C. \& Eilek, J.~A.} 2003,
{Hankins, T.~H. \etal\ } 2003,
\textit{Nature}, 422, 141.

%\bibitem[Johnson \& Gwin (2012)]{jg12}
%{Johnson, M.~D. \& Gwinn, C.~R.} 2012
%\textit{ApJ}, 755, 179.

%\bibitem[Johnson \etal\  (2012)]{jgd12}
%{Johnson, M.~D., Gwinn, C.~R. \& Demorest, P.} 2012
%\textit{ApJ}, 758, 8.

\bibitem[Johnston \etal\  (2001)]{jvkb01}
%{Johnston, S., van Straten, W., Kramer, M. \& Bailes, M.} 2001,
{Johnston, S. \etal\ } 2001,
\textit{ApJ}, 549, L101.

\bibitem[Keane \etal\  (2011)]{kkl+11} 
%{Keane, E.~F., Kramer, M., Lyne, A.~G., Stappers, B.~W. \& McLaughlin, M.~A.} 2011, 
{Keane, E.~F. \etal\ } 2011,
\textit{MNRAS}, 415, 3065.

\bibitem[Keane (2010a)]{kea10a}
{Keane, E.~F.} 2010,
``Transient Radio Neutron Stars'', Proceedings of HTRA-IV. May 5 - 7, 2010. Agios Nikolaos, Crete Greece. 

\bibitem[Keane (2010b)]{kea10b}
{Keane, E.~F.} 2010,
``The Transient Radio Sky'', PhD thesis, University of Manchester.

\bibitem[Keane \& McLaughlin (2011)]{km11}
{Keane, E.~F. \& McLaughlin, M.~A.} 2011,
\textit{Bulletin of the Astronomical Society of India}, 39, 1.

%\bibitem[Kocz \etal\  (2011)]{kbb+11}
%{Kocz, J., Bailes, M., Barnes, D. Burke-Spolaor, S. \& Levin, L.} 2011,
%\textit{MNRAS}, in press.

\bibitem[Kramer \etal\  (2006)]{klo+06}
%{Kramer, M., Lyne, A.~G., O'Brien, J.~T., Jordan, C.~A. \& Lorimer, D.~R.} 2006,
{Kramer, M. \etal\ } 2006,
\textit{Science}, 312, 549.

\bibitem[Liu \etal\  (2012)]{lkl+12}
%{Liu, K., Keane, E.~F., Lee, K.~J., Kramer, M., Cordes, J.~M. \& Purver, M.~B.} 2012,
{Liu, K. \etal\ } 2012,
\textit{MNRAS}, 420, 361.

%\bibitem[Lorimer \etal\  (2007)]{lbm+11}
%{Lorimer, D.~R., Bailes, M., McLaughlin, M.~A., Narkevic, D.~J. \& Crawford, F.} 2007,
%\textit{Science}, 318, 777.

\bibitem[Lorimer \etal\  (2012)]{llm+12}
%{Lorimer, D.~R., Lyne, A.~G., McLaughlin, M.~A., Pavlov, G.~G. \& Chang, C.} 2012,
{Lorimer, D.~R. \etal\ } 2012,
\textit{ApJ, submitted}, astro-ph/1208.6576.

%\bibitem[Lyne \etal\  (2009)]{lmk+09}
%%{Lyne, A.~G., McLaughlin, M.~A., Keane, E.~F., Kramer, M., Espinoza, C.~M., Stappers, B.~W., Palliyaguru, N.~T. \& Miller, J.} 2009,
%{Lyne, A.~G. \etal\ } 2009,
%\textit{MNRAS}, 400, 1439.

\bibitem[Lyne \etal\  (2010)]{lhk+10}
%{Lyne, A.~G., Hobbs, G., Kramer, M., Stairs, I. \& Stappers, B.~W.} 2010,
{Lyne, A.~G. \etal\ } 2010,
\textit{Science}, 329, 408.

\bibitem[Lyutikov \etal\  (1999)]{lbm99}
{Lyutikov, M., Blandford, R. D. \& Machabeli, G.} 1999,
\textit{MNRAS}, 305, 338.

\bibitem[Maron \etal\  (2000)]{mkk+00}
%{Maron, O., Kijak, J., Kramer, M., \& Wielebinski, R.} 2000,
{Maron, O. \etal\ } 2000,
\textit{A\&AS}, 147, 195.

\bibitem[Melrose (2004)]{mel04b}
{Melrose, D.} 2004,
in Young Neutron Stars and Their Environments, Vol. 1, 
IAU Symposium 218, Astronomical Society of the Pacific, 
San Francisco, 349.

\bibitem[Michel (1973)]{mic73}
{Michel F. C.} 1973, 
\textit{ApJ}, 180, L133.

\bibitem[Rickett (1975)]{ric75}
{Rickett, B.~J.} 1975,
\textit{ApJ}, 197, 185.

\bibitem[Ruderman \& Sutherland (1975)]{rs75}
{Ruderman, M. A. \& Sutherland, P. G.} 1975,
\textit{ApJ}, 196, 51.

\bibitem[Seymour \& Lorimer (2012)]{sl12}
{Seymour, A.~D. \& Lorimer, D.~R.} 2012,
\textit{MNRAS, in press}, astro-ph/1209.5645

\bibitem[Spitkovsky (2006)]{spi06}
{Spitkovsky, A.} 2006,
\textit{ApJ}, 648, L51.

\bibitem[Timokhin (2010)]{tim10}
{Timokhin, A.~N.} 2010,
\textit{MNRAS}, 408, L41.

\bibitem[Weltevrede \etal\  (2006)]{wes06}
{Weltevrede, P., Edwards, R.~T. \& Stappers, B.~W.} 2006,
\textit{A\&A}, 445, 243.

%\bibitem[Weltevrede \etal\  (2006)]{wsrw06}
%{Weltevrede, P., Stappers, B. W., Rankin, J. M. \& Wright, G. A. E.} 2006,
%\textit{ApJ}, 645, L149.

%\bibitem[Weltevrede \etal\  (2011)]{wje11}
%{Weltevrede, P., Johnston, S. \& Espinoza, C.~M.} 2011,
%\textit{MNRAS}, 411, 1917.

\bibitem[Young \etal\  (2012)]{ysw+12}
%{Young, N.~J., Stappers, B.~W., Weltevrede, P., Lyne, A.~G. \& Kramer, M.} 2012,
{Young, N.~J. \etal\ } 2012,
\textit{MNRAS, in press}, astro-ph/1208.3868.


\end{thebibliography}
\end{document}